\documentclass[twocolumn,a4paper,superscriptaddress,nofootinbib,showpacs,pra]{revtex4}

\usepackage{amsfonts,amsmath,amsthm,graphicx}

\newtheorem{theorem}{Theorem}
\newtheorem*{thmnonum}{Theorem}
\newtheorem*{lemma}{Lemma}
\newtheorem{result}[theorem]{Result}

\begin{document}

\title{Entanglement under restricted operations: Analogy to
  mixed-state entanglement}
\author{Stephen D. Bartlett}
\affiliation{School of Physics, The University of Sydney,
  New South Wales 2006, Australia}
\author{Andrew C. Doherty}
\affiliation{School of Physical Sciences, The University of Queensland,
  Queensland 4072, Australia}
\author{Robert W. Spekkens}
\affiliation{Perimeter Institute for Theoretical Physics, 31
  Caroline St.~N., Waterloo, Ontario N2J 2Y5, Canada}
\author{H. M. Wiseman}
\affiliation{Centre for Quantum Computer Technology, Centre for
  Quantum Dynamics, School of Science, Griffith University, Brisbane,
  4111 Australia}
\date{3 February 2006}

\begin{abstract}
We show that the classification of bi-partite pure entangled states
when local quantum operations are restricted yields a structure that
is analogous in many respects to that of mixed-state entanglement.
Specifically, we develop this analogy by restricting operations
through local superselection rules, and show that such exotic
phenomena as bound entanglement and activation arise using pure
states in this setting.  This analogy aids in resolving several
conceptual puzzles in the study of entanglement under restricted
operations.  In particular, we demonstrate that several types of
quantum optical states that possess confusing entanglement
properties are analogous to bound entangled states.  Also, the
classification of pure-state entanglement under restricted
operations can be much simpler than for mixed-state entanglement.
For instance, in the case of local Abelian superselection rules all
questions concerning distillability can be resolved.
\end{abstract}


\maketitle

\section{Introduction}

Entanglement of quantum systems is a potentially powerful resource
for quantum information processing~\cite{Nie00}.  However, in the
presence of noise, it is currently not known precisely \emph{which}
entangled states are useful, and a vast theory of mixed-state
entanglement has developed to classify states according to their
entanglement properties~\cite{Hor01}.

In this paper, we show that the theory of pure-state entanglement
when quantum operations are restricted -- described formally by a
superselection rule -- precisely replicates the structure of
mixed-state entanglement, including such exotic properties as bound
entanglement and activation.  This analogy is useful both for the
theory of mixed-state entanglement, and for that of pure-state
entanglement under restricted operations. After over a decade of
debate on issues such as the non-locality of a single
photon~\cite{Tan91,Har94,GHZ95,Har95,Enk05b,Hes04,Bab04} and the
role of a phase reference in quantum
teleportation~\cite{Rud01,Enk02,Rud01b,vEF02b,Nem02,Hey03,San03,Wis03c,Wis04},
we resolve these conceptual issues by demonstrating that
entanglement under restricted operations can be viewed as
\emph{bound} by the restriction. In addition, we demonstrate that
the surprising results for entanglement under
constraints~\cite{Ver03,Wis03,BW03,Sch04a,Sch04b} arise from the
coexistence of two distinct operational notions of entanglement, and
that distinguishing these notions realizes the entire structure of
the pre-existing mixed-state entanglement theory.  Thus, we
demonstrate that the specialized concepts of the field of
mixed-state entanglement (such as activation and multi-copy
distillation) can be applied to a wide variety of practical
situations.  Moreover, unsolved questions for mixed-state
entanglement have analogous questions in the context of pure-state
entanglement under restrictions, and these can be answered in some
cases.  It is hoped that this formal analogue of the complex and
surprising structure of mixed-state entanglement in another
situation -- one that is conceptually straightforward to understand
and interpret -- will ultimately lead to new results in mixed-state
entanglement theory.

\section{Classifying Mixed-State Entanglement}
\label{sec:MSE}

In this section, we present some known results for the
classification of mixed-state entanglement, with our own bias and
some new terminology.  For an extensive review of mixed-state
entanglement, see~\cite{Hor01}.

Central to the theory of entanglement is the classification of the
states of a quantum system shared between two parties (Alice and
Bob) who can perform only local quantum operations and classical
communication (LOCC).  This limitation on their operations means on
the one hand that certain states cannot be prepared by the two
parties starting from some uncorrelated fiducial state, and on the
other hand that certain states shared by the two parties may serve
as resources allowing them to perform tasks not possible with LOCC
alone.  In this paper, it will be important to distinguish between
various sets of states characterized by either \emph{(i)} the
operations required to prepare them, or \emph{(ii)} the resource
they provide for quantum information processing tasks.  To emphasise
the distinctions between these sets, we adopt a slightly
unconventional terminology for mixed-state entanglement.  First, we
identify the class of bi-partite states that are \emph{locally
preparable}, that is, preparable by LOCC (starting with some
uncorrelated fiducial state).  We denote this class LP.  Second, we
identify the class of states that are
\emph{distillable}~\cite{Ben96}, denoted D. States are distillable
if $n$ copies can be converted into $nr$ pure maximally entangled
states via LOCC for some $r>0$ in the limit $n\to\infty$.

A \emph{pure} state is either locally preparable or distillable
(either in LP or in D), depending on whether it is a product state
or not (i.e., a state of the form $|\psi\rangle_A \otimes
|\phi\rangle_B$ or not).  For mixed states, the set LP is the set of
states that possess a convex decomposition into product states (the
\emph{separable} states).  Identifying the class of mixed states
that are distillable is important for quantum information
processing, but unfortunately it is not known how to determine if a
general bipartite mixed state is distillable or not~\cite{Hor01}.
One property of the class D is certain, though: in contrast to the
situation for pure states, there are mixed states that are neither
locally preparable nor distillable, called \emph{bound entangled}
states~\cite{Hor98}.

Part of the difficulty in identifying the set of distillable mixed
states arises from the asymptotic nature of the definition of
distillability, as it is not known how to characterize all possible
distillation protocols that act on a potentially infinite number of
copies.  In the following, we will make use of a related class with
a simpler characterisation: the class of states that are
\emph{1-distillable}~\cite{Div00,Dur00}, denoted 1-D.  We define and
motivate this class as follows.  First, we note that distillability
is decidable on a $2{\times}2$-dimensional space, wherein all
separable states are in LP, and all non-separable states are
distillable~\cite{Hor97}.  On an arbitrary bi-partite space, we
define a state $\rho$ to be 1-distillable if there exists an
operation implementable with LOCC, represented by a completely
positive map $\mathcal{E}$~\cite{Nie00}, that maps $\rho$ onto a
$2{\times}2$-dimensional subspace of the bi-partite system such that
$\mathcal{E}(\rho)$ is non-separable (and thus distillable).  If a
state is 1-distillable then it is distillable.  For pure states,
1-distillability is equivalent to distillability, and thus every
pure state is either locally preparable or 1-distillable.  This is
not the case for mixed states. Due to the existence of bound
entangled states (i.e., states that are neither locally preparable
nor distillable), and the fact that 1-D $\subset$ D, there exist
mixed states that are neither locally preparable \emph{nor}
1-distillable. We shall refer to all such states as \emph{1-bound}.

\begin{figure}
\includegraphics[width=3.25in]{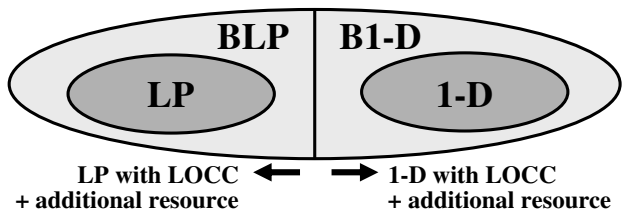}
\caption{Illustration of the division of all bi-partite mixed states
  into four classes.  When restricted to LOCC, there is a proper gap
  between what is locally preparable and what is 1-distillable.  This
  gap contains 1-bound states.  If an additional resource is supplied,
  allowing for all PPT-preserving operations, then all 1-bound states
  either become locally preparable (BLP) or become 1-distillable (B1-D).}
\label{fig:Fig1}
\end{figure}

Remarkably, by appropriately extending the set of operations that
Alice and Bob can perform beyond LOCC, all states become either
locally preparable or 1-distillable.  We describe this extension of
operations as supplementing LOCC with an additional \emph{resource}.
Clearly, additional power will affect the boundaries of what Alice
and Bob can prepare or distill; we are interested in a resource that
precisely removes the proper gap between LP and 1-D. Consider
extending LOCC to allow all operations that preserve the positivity
of the partial transpose of states~\cite{Rai01}. With this
additional resource, all states with positive partial transpose
(PPT) can be prepared locally.  All states that are not PPT are
1-distillable with this additional power in the sense that they can
be mapped by a PPT-preserving operation $\mathcal{E}$ onto a
$2{\times}2$-dimensional space such that $\mathcal{E}(\rho)$ is
non-separable~\cite{Egg01}.  States that are not locally preparable
with LOCC, but locally preparable given LOCC plus the additional
resource, can be said to \emph{become} locally preparable given the
resource, denoted BLP.  Similarly, the class of states that are not
1-distillable but become 1-distillable given the resource we denote
as B1-D. For mixed bi-partite states under PPT-preserving
operations, the class BLP contains all PPT bound entangled states
and the class B1-D contains all non-PPT states that are not
1-distillable; both classes are non-empty~\cite{Hor98,Div00,Dur00}.
See Fig.~\ref{fig:Fig1}.

\begin{figure}
\includegraphics[width=3.25in]{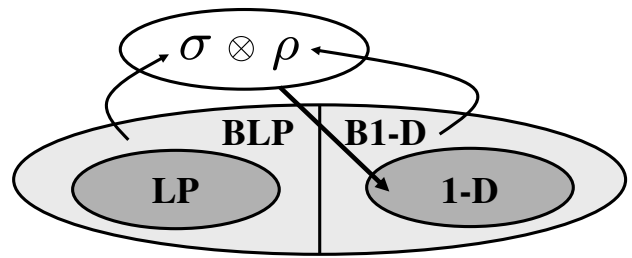}
\caption{Illustration of activation of bound entanglement.  A BLP
state $\sigma$ can be used to ``activate'' the entanglement of a
B1-D state $\rho$, i.e., the combined state $\rho \otimes \sigma$ is
1-distillable.} \label{fig:Fig2}
\end{figure}

The categories BLP and B1-D are related in an interesting way.
Through an isomorphism between bi-partite quantum states and quantum
operations, any PPT-preserving operation can be implemented
probabilistically using LOCC and a specific state in BLP (i.e., a
specific PPT bound entangled state)~\cite{Cir01}.  Recall that any
B1-D state becomes 1-distillable if Alice and Bob are given the
additional resource of all PPT-preserving operations.  Thus, for
every $\rho \in \text{B1-D}$ there exists a state $\sigma \in
\text{BLP}$ such that $\sigma \otimes \rho$ is 1-distillable.  We
say that the state $\sigma \in \text{BLP}$ \emph{activates} the
entanglement of the state $\rho \in \text{B1-D}$ using only LOCC
operations~\cite{Hor99}; see Fig.~\ref{fig:Fig2}.

Another remarkable feature of mixed-state entanglement is that,
although states in B1-D are not 1-distillable, they may nevertheless
be distillable~\cite{Wat04}.  We define a state $\rho$ to be
$n$-distillable (in $n$-D) if there exists an LOCC operation
$\mathcal{E}_n$ onto a $2\times 2$-dimensional space such that
$\mathcal{E}_n(\rho^{\otimes n})$ is non-separable.  In other words,
the joint state $\rho^{\otimes n}$ is 1-distillable.  If a state is
$n$-distillable for some $n$ then it is distillable.  (In fact, it
has been shown~\cite{Wat04} that $n$-D is a proper subset of D for
all finite $n$.)  Thus, mixed-state entanglement exhibits multi-copy
distillability, meaning that there exist states in B1-D that are not
1-distillable but that are $n$-distillable for some $n\geq 2$.

There remain, however, many open questions regarding the general
structure of mixed-state entanglement.  Perhaps the most important
question from the point of view of quantum information processing
is: Are \emph{all} states in B1-D distillable?

\section{An analogy in quantum optics}
\label{sec:optics}

To introduce the concepts and results developed later in this paper,
we first begin by providing a simple example of how the phenomena
arising in the context of mixed-state entanglement have precise
analogues in the structure of pure-state entanglement under a
restriction on operations. Specifically, we consider some
well-studied states from quantum optics and the restriction of a
\emph{local photon-number superselection rule}.

\subsection{Local photon-number superselection rule}

The restriction of a local photon-number superselection rule implies
that a party cannot prepare coherent superpositions of states of
different local photon number (starting with states without
coherence), nor measure such coherences, nor implement a
transformation that creates such coherence.  For instance, if Alice
is restricted by a local photon-number superselection rule, she
cannot prepare a state of the form $\frac{1}{\sqrt{2}}(|0\rangle_A +
|1\rangle_A)$, where $|n\rangle_A$ denotes an $n$-photon eigenstate
of a mode in Alice's possession.  However, she can prepare the state
$\frac{1}{\sqrt{2}}(|01\rangle_A + |10\rangle_A)$ on a pair of modes
in her possession (where $|01\rangle_A =
|0\rangle_{A_1}\otimes|1\rangle_{A_2}$, etc.), because this state is
an eigenstate of total local photon number.

A local photon-number superselection rule applies to multi-party
quantum optics experiments when the parties do not share a common
phase reference~\cite{Ver03,San03}.  This connection between
superselection rules and reference frames can be seen as follows. In
optical experiments, states of an optical mode are always referred
to some phase reference.  Consider several optical modes distributed
between two parties, Alice and Bob.  Suppose there is a third party,
Charlie, who has a local phase reference -- for example, a high
intensity laser -- to which the quantum states of Alice and Bob's
optical modes can be referred. Suppose further that Alice and Bob do
not share this phase reference, i.e., their lasers are not
phase-locked with Charlie's.  The relative phase between their phase
references and Charlie's is therefore completely unknown.

We now demonstrate that this unknown phase relation leads to a local
photon-number superselection rule.  Let Alice prepare a quantum
state of her local optical modes, which she represents by a density
operator $\rho_A$ relative to her phase reference.  If the relative
phase between Alice and Charlie's phase references was known to be
$\phi$, then this same state would be represented by the density
operator $e^{-i\phi \hat{N}_A} \rho_A e^{i\phi \hat{N}_A}$ relative
to Charlie's phase reference, where $\hat{N}_A$ is Alice's local
photon number operator.  Given that $\phi$ is completely unknown,
one must average over its possible values to obtain the state
relative to Charlie. This state is
\begin{equation}
  \label{eq:LPNSSR1}
  \mathcal{U}_A[\rho_A] \equiv \int_0^{2\pi} \frac{d\phi}{2\pi}\,
  e^{-i\phi \hat{N}_A} \rho_A e^{i\phi \hat{N}_A} \,,
\end{equation}
which is equivalent to
\begin{equation}
  \label{eq:LPNSSR2}
  \mathcal{U}_A[\rho_A] = \sum_n \Pi^A_n \rho_A \Pi^A_n \,,
\end{equation}
where $\Pi^A_n$ is the projector onto the $n$th eigenspace of
$\hat{N}_A$.  The map $\mathcal{U}_A$ removes all coherence between
states of differing total photon number on Alice's systems.
Similarly, any operations Alice implements relative to her phase
reference are redescribed relative to Charlie's phase reference as
operations that commute with $\mathcal{U}_A$.  Thus, relative to
Charlie, Alice experiences a restriction on operations that is
described by a superselection rule for local photon number
$\hat{N}_A$ as defined in~\cite{BW03}.  A similar argument applies
to states and operations of Bob relative to Charlie, characterised
by a map $\mathcal{U}_B$. Thus, the situation where Alice and Bob
lack Charlie's phase reference is a restriction formally equivalent
to a superselection rule for local photon number.

We note that although the term ``superselection rule'' was initially
introduced to describe an \emph{in principle} restriction on quantum
states and operations~\cite{WWW52}, it has been emphasized by
Aharonov and Susskind~\cite{Aha67} that whether or not coherent
superpositions of a particular observable are possible is a
practical matter, depending on the availability of a suitable
reference system.  Modern arguments in favour of this view may be
found in Refs.~\cite{BW03,KMP04,BRS05}, and we follow the practice
of using the term ``superselection rule'' to describe both in
principle and practical restrictions on operations.

\subsection{Bound entanglement in pure-state quantum optics}

In such situations, there has been considerable debate over the
entanglement properties of certain types of states, such as the
two-mode single-photon state~\cite{Tan91,Har94,GHZ95,Har95,Enk05b}
\begin{equation}
  \label{V-EPR}
  \tfrac{1}{\sqrt{2}} (|0\rangle_A |1\rangle_B +
  |1\rangle_A |0\rangle_B)\,.
\end{equation}
There is a temptation to say that this state is entangled simply
because of its nonproduct form.  However, it is far more useful to
consider whether or not this state satisfies certain
\emph{operational} notions of entanglement.  One such notion is
whether a state can be used to violate a Bell inequality.  Another
is whether it is useful as a resource for quantum information
processing, for instance, to teleport qubits or implement a dense
coding protocol.  In the context of a local photon-number
superselection rule, this two-mode single-photon state fails to
satisfy either of these notions of entanglement, because all such
tasks would require Alice and Bob to violate the local photon-number
superselection rule.  A different but equally common notion of
entanglement is that a state is entangled if it cannot be prepared
by LOCC.  The two-mode single-photon state certainly \emph{does} fit
\emph{this} notion because the pure nonproduct states cannot be
prepared by LOCC. Thus we see that operational notions of
entanglement that coincided for pure states under unrestricted LOCC,
namely being not locally preparable and being useful as a resource
for tasks such as teleportation or violating a Bell inequality, do
not coincide under a local photon-number superselection rule, and
the state in question is judged entangled by one notion and not the
other.\footnote{Of course, if there is no local photon-number
superselection rule, this state would satisfy all of these notions
of entanglement, as emphasised by van Enk~\cite{Enk05b}.  In
particular, no such superselection rule would apply if all parties
share a common phase reference, as discussed in
Sec.~\ref{subsec:phaseref}.}

This already has the flavour of the phenomenon of bound
entanglement. However, strictly speaking, the two notions of
entanglement that were explored in Sec.~\ref{sec:MSE} were local
preparability and \emph{1-distillability}. As one might expect from
the above comments, these two notions do not coincide either, as we
now show.

Consider a state of the form
\begin{equation}
  \label{E-EPR} \tfrac{1}{\sqrt{2}} (|01\rangle_A |10\rangle_B +
  |10\rangle_A |01\rangle_B)\,.
\end{equation}
This state is certainly not locally preparable.  In addition, it
\emph{can} be used to violate a Bell inequality, implement dense
coding, and so on, despite the superselection rule.  This is because
Alice and Bob can still implement any measurements they please in
the 2-dimensional subspaces spanned by $|01\rangle$ and
$|10\rangle$. Thus, a useful notion of distillability for a
bi-partite pure state in the context of a local photon number
superselection rule is whether $n$ copies of the state can be
converted into $nr$ copies of $\tfrac{1}{\sqrt{2}} (|01\rangle_A
|10\rangle_B + |10\rangle_A |01\rangle_B)$ for some $r>0$.  A useful
notion of \emph{1-distillability} for a bi-partite pure state in the
context of a local photon number superselection rule is whether it
can be projected to a nonproduct state in some $2\times 2$ subspace,
where the 2-dimensional local spaces are \emph{eigenspaces of local
photon number}.

By this definition, the state $\tfrac{1}{\sqrt{2}} (|0\rangle_A
|1\rangle_B + |1\rangle_A |0\rangle_B)$ is clearly not 1-distillable
under the local photon number superselection rule because the
subspace spanned by $\{ |0\rangle,|1\rangle \}$ cannot be mapped to
the subspace spanned by $\{ |01\rangle,|10\rangle \}$ under the
restricted operations.  This establishes the existence of pure
states that are neither locally preparable nor 1-distillable under
the local photon number superselection rule.  Thus, they are
analogous to the \emph{1-bound} states introduced for mixed-state
entanglement.



Another class of states whose entanglement properties have been
discussed recently in the quantum optics literature are those that
are separable but not locally preparable under a local photon-number
superselection rule~\cite{Rud01,Ver03}. Examples of such
states\footnote{Refs.~\cite{Rud01,Ver03} considered
  states such as the equal mixture of
  $|{+}\rangle_A|{+}\rangle_B$ and $|{-}\rangle_A|{-}\rangle_B$.
  For simplicity, we restrict our attention to pure states.} are
\begin{equation}
\label{refbits}
  |{+}\rangle_A|{+}\rangle_B\,, \qquad |{-}\rangle_A|{-}\rangle_B\,,
\end{equation}
where $|\pm\rangle = \frac{1}{\sqrt{2}} (|0\rangle\pm|1\rangle)$.
Because of the superselection rule, these states cannot be prepared
locally.  But they are not 1-distillable either because they are
product states.  Thus, they also lie in the gap between what is
locally preparable and what is 1-distillable. Verstraete and
Cirac~\cite{Ver03} identified such states as a ``new type of
nonlocal resource'', and van Enk~\cite{Enk05} identified states of
the form of Eq.~(\ref{refbits}) as a standard unit of this nonlocal
resource, which he called a ``refbit''.  We can identify these
states also as analogues of the 1-bound states of mixed-state
entanglement.

\subsection{A resource to ``lift'' the superselection rule}
\label{subsec:phaseref}

In the context of a superselection rule, there is also a resource
that precisely removes the gap between what is locally preparable
and 1-distillable (as occurred in mixed-state entanglement by
extending LOCC to all PPT-preserving operations).  Recall, as
described above, that a local photon-number superselection rule
applies if Alice and Bob are uncorrelated with the phase reference
of Charlie, who is preparing the bi-partite quantum states.
Clearly, if Alice and Bob are given phase references that are
precisely correlated with Charlie's, then they no longer face any
restrictions beyond that of LOCC. Thus, for Alice and Bob to possess
a shared phase reference is for them to possess a resource that
``lifts'' the superselection rule. Given this resource, states such
as $\tfrac{1}{\sqrt{2}} (|0\rangle_A |1\rangle_B + |1\rangle_A
|0\rangle_B)$ \emph{become} 1-distillable, while states such as
$|+\rangle_A |+\rangle_B$ \emph{become} locally preparable.


\subsection{Activation and distillation in pure-state quantum
optics}

Finally, we demonstrate that there exist analogous processes of
activation and multi-copy distillation in this scenario.  Both of
these processes have been discussed (albeit using different
terminology) by van Enk~\cite{Enk05} for the specific quantum
optical state examples we present here.

Combining $\tfrac{1}{\sqrt{2}} (|0\rangle_{A} |1\rangle_{B} +
|1\rangle_{A} |0\rangle_{B})$ (a state which, by itself, is
\emph{not} 1-distillable under a local photon-number superselection
rule) with $|{+}\rangle_{A} |{+}\rangle_{B}$ one obtains a state
that \emph{is} 1-distillable.  The state $|{+}\rangle_{A}
|{+}\rangle_{B}$ is said to \emph{activate} the entanglement of
$\tfrac{1}{\sqrt{2}} (|0\rangle_{A} |1\rangle_{B} + |1\rangle_{A}
|0\rangle_{B})$.  This is seen as follows.  Let Alice and Bob both
perform a quantum non-demolition measurement of local photon number,
and post-select the case where they both find a local photon number
of one.  The resulting state is
\begin{multline}
    \Pi^A_1 \otimes \Pi^B_1 [ \tfrac{1}{\sqrt{2}}(|0\rangle_{A} |1\rangle_{B}
+ |1\rangle_{A} |0\rangle_{B}) |{+}\rangle_{A}
|{+}\rangle_{B} ] \\
    \propto \tfrac{1}{\sqrt{2}}( |01\rangle_{A} |10\rangle_B +
|10\rangle_A
    |01\rangle_B) \,.
\end{multline}

We note that the controversy over the use of the state
$\tfrac{1}{\sqrt{2}} (|0\rangle_{A} |1\rangle_{B} + |1\rangle_{A}
|0\rangle_{B})$ to demonstrate quantum
nonlocality~\cite{Tan91,Har94,GHZ95,Har95,Hes04,Bab04} can be
resolved by recognizing the role of activation. As we have shown,
this state is not 1-distillable when Alice and Bob do not share a
correlated local phase reference (i.e., when a local photon-number
superselection rule applies). However, violations of a Bell
inequality has recently been demonstrated experimentally using this
state~\cite{Hes04,Bab04}.  One can take two different perspectives
on such an experiment.  It is illustrative to consider them both.

In Ref.~\cite{Hes04}, in addition to the state $\tfrac{1}{\sqrt{2}}
(|0\rangle_{A} |1\rangle_{B} + |1\rangle_{A} |0\rangle_{B})$, a
correlated pair of coherent states $|\alpha\rangle_A
|\alpha\rangle_B$, where $|\alpha\rangle \equiv \sum_n (
e^{-|\alpha|^2/2} \alpha^n / \sqrt{n!}) |n\rangle$, are assumed to
be shared between Alice and Bob. These modes are used as the local
oscillators in the homodyne detections at each wing.  Noting that
neither $\tfrac{1}{\sqrt{2}} (|0\rangle_{A} |1\rangle_{B} +
|1\rangle_{A} |0\rangle_{B})$ nor $|\alpha\rangle_A
|\alpha\rangle_B$ are 1-distillable under the superselection rule,
it is unclear how it is possible to violate the Bell inequality
using such resources.  The resolution of the puzzle is that the
product of coherent states $|\alpha\rangle_A |\alpha\rangle_B$ (like
the state $|{+}\rangle_A |{+}\rangle_B$) \emph{activates} the
entanglement of the two-mode single photon state.  To see this, we
note that the same measurement of local photon number as described
above projects the state onto a non-product state of random but
definite local photon number, allowing for a demonstration of
nonlocality within the constraints of the superselection rule. (Such
a measurement is, in fact, implemented using an ideal homodyne
detection. Loosely speaking, each observer's homodyne detection
apparatus couples the two local modes at a beam splitter and then
measures the number of photons in each of the two output ports. This
incorporates a measurement of the total local photon number because
the latter quantity can be obtained as the sum of the number in each
output port. The difference of these two photocounts, which is
typically the quantity of interest in homodyne detection, yields the
information necessary to demonstrate the Bell inequality violation.)

An experimental demonstration of nonlocality using the two-mode
single photon state can also be described as follows~\cite{Bab04}.
Rather than treating the local oscillators as coherent states, they
are treated as correlated classical phase references. In this case,
they constitute an additional resource that ``lifts'' the
restriction of the local photon-number superselection rule, and the
state $\tfrac{1}{\sqrt{2}} (|0\rangle_{A} |1\rangle_{B} +
|1\rangle_{A} |0\rangle_{B})$ \emph{becomes} 1-distillable. These
two alternative descriptions are equally valid~\cite{BRS05}.

The existence of such activation processes also resolves a
controversy concerning the source of entanglement in the
experimental realization of continuous-variable quantum
teleportation~\cite{Fur98}.  Again, we consider two different
perspectives on the experiment.

The first perspective is a variant of the one presented by Rudolph
and Sanders~\cite{Rud01}.  In our language, it can be synopsized as
follows.  Alice and Bob are presumed to be restricted in the
operations they can perform by a local photon-number superselection
rule.  They share a two-mode squeezed state $|\gamma\rangle
=\sqrt{1-\gamma^{2}}\sum_{n=0}^{\infty}\gamma^{n}|n,n\rangle$ where
$0\leq\gamma\leq 1$.  In addition, they share two other modes
prepared in a product of coherent states $|\alpha\rangle
|\alpha\rangle$.\footnote{The state assigned to this pair of
resources in Ref.~\cite{Rud01} is simply a mixed version of the one
we consider here.}  The former is the purported entanglement
resource in the teleportation protocol, while the latter is a
quantum version of a shared phase reference.  These states are
analogous to $\tfrac{1}{\sqrt{2}} (|0\rangle_{A} |1\rangle_{B} +
|1\rangle_{A} |0\rangle_{B})$ and $|{+}\rangle_{A} |{+}\rangle_{B}$
respectively -- neither is 1-distillable when considered on its own.
So the question arises as to how teleportation could possibly have
been achieved when neither the purported entanglement resource nor
the quantum shared phase reference are 1-distillable. The resolution
to this puzzle is that although individually, neither is
1-distillable, together they are: the quantum shared phase reference
\emph{activates} the entanglement in the two-mode squeezed state.

The second perspective is one wherein the shared phase reference is
treated classically~\cite{Fur98}.  As described above, this acts as
a resource which lifts the superselection rule, and causes the
two-mode squeezed state to become 1-distillable.

An analogue of multi-copy distillation can also be demonstrated in
our quantum optical example. For instance, the state
$\tfrac{1}{\sqrt{2}} (|0\rangle_{A} |1\rangle_{B} + |1\rangle_{A}
|0\rangle_{B})$ is 2-distillable. The protocol, introduced in
Ref.~\cite{Wis03} and discussed in greater detail in
Ref.~\cite{Ans04}, is as follows.  As in the activation example
above, Alice and Bob both perform a quantum non-demolition
measurement of local photon number (on both local modes) and
post-select the case where they both find a local photon number of
one.  The resulting state is
\begin{multline}
    \Pi^A_1 \otimes \Pi^B_1 [\tfrac{1}{\sqrt{2}}(|0\rangle_{A}
    |1\rangle_{B}
    + |1\rangle_{A} |0\rangle_{B})]^{\otimes 2} \\
    \propto \tfrac{1}{\sqrt{2}}( |01\rangle_{A} |10\rangle_B +
    |10\rangle_A |01\rangle_B) \,,
\end{multline}
where $|\psi\rangle^{\otimes 2} = |\psi\rangle |\psi\rangle$. A
process very similar to this 2-copy distillation has been
demonstrated in quantum optics experiments (c.f.~\cite{Shi88}),
where correlated but unentangled photon pairs from parametric
downconversion were made incident on the two input modes of a
beamsplitter, so each photon transforms to a state of the form
$\tfrac{1}{\sqrt{2}} (|0\rangle_{A} |1\rangle_{B} + |1\rangle_{A}
|0\rangle_{B})$.  Subsequently, measurements on the two output modes
are postselected for one photon detection at each output mode. The
fact that their postselected results are consistent with a
description of an entangled state demonstrates that the entanglement
of the state $\tfrac{1}{\sqrt{2}} (|0\rangle_{A} |1\rangle_{B} +
|1\rangle_{A} |0\rangle_{B})$ has been distilled by making use of
two copies.

We see that the remarkable (and often confusing) entanglement
properties of states when local operations are restricted can be
understood by recognizing that different operational notions of
entanglement do not coincide in this case, leaving a structure akin
to that of mixed-state entanglement.

\section{Pure-state entanglement under general restrictions}

We now develop the analogy between mixed-state entanglement and
pure-state entanglement when the allowed local quantum operations
are restricted by a \emph{general} (not necessarily Abelian)
superselection rule. We continue to consider only pure states,
because, although one could characterise mixed-state entanglement
under such restrictions, the classification of such states would be
at least as difficult as unrestricted mixed-state entanglement.

\subsection{Restricting operations through general superselection
rules}

We formulate a restriction on operations generally in the form of a
superselection rule (SSR) associated with a finite or compact Lie
group $G$~\cite{BW03,KMP04}. (A different concept of entanglement
under restrictions on operations is discussed in~\cite{Barnum03}.)

The superselection rule we describe can be defined operationally as
follows. Suppose Alice and Bob share a pair of systems, described by
a Hilbert space $\mathcal{H}^A \otimes \mathcal{H}^B$, the states on
which were prepared and described by a third party, Charlie. Suppose
further that the local reference frames of Alice, Bob and Charlie,
which transform via a group $G$, are uncorrelated: that is, the
element $g\in G$ relating Alice's and Charlie's local frames is
completely unknown, as is the element $g'\in G$ relating Bob and
Charlie's local frames. It follows that a preparation represented by
a density matrix $\rho$ on $\mathcal{H}^A$ relative to Alice's frame
is represented by the density matrix $\mathcal{G}_{A}[\rho]$
relative to Charlie's frame, where
\begin{equation}
  \label{eq:AveragedState}
  \mathcal{G}_A[\rho] \equiv \int_G \text{d}v(g)\,
  T^{A}(g) \rho T^{A \dag}(g) \, ,
\end{equation}
with $T^A(g)$ a unitary representation of $g$ on $\mathcal{H}^A$,
and $\text{d}v$ the group-invariant (Haar) measure.  The operations
that Alice can implement relative to Charlie's frame are represented
by completely positive maps $\mathcal{O}_A$ that commute with
$\mathcal{G}_{A}$.  A similar result holds for the operations that
Bob can implement. The joint LOCC operations that Alice and Bob can
implement relative to Charlie's frame are those represented by maps
$\mathcal{O}_{AB}$ that commute with $\mathcal{G}_{A} \otimes
\mathcal{G}_{B}$.  These are said to be locally
$G$-invariant~\cite{BW03}.  This restriction on operations is
referred to as a local superselection rule for $G$.

A local superselection rule for $G$ induces the following structure
in the local Hilbert spaces (we consider $\mathcal{H}^A$):
\begin{equation}
    \mathcal{H}^A = \bigoplus_n \mathcal{H}^A_n \,,
\end{equation}
i.e., each local Hilbert space is split into ``charge sectors''
labeled by $n$ and each carrying inequivalent representations
$T^A_n$ of $G$.  Each sector can be further decomposed into a tensor
product,
\begin{equation}
  \mathcal{H}^A_n = \mathcal{M}^A_{n} \otimes \mathcal{N}^A_n \,,
\end{equation}
of a subsystem $\mathcal{M}^A_n$ carrying an irreducible
representation $T^A_n$ and a subsystem $\mathcal{N}^A_n$ carrying a
trivial representation of $G$.  For an Abelian superselection rule,
such as the photon-number superselection rule discussed in
Sec.~\ref{sec:optics}, the subsystems $\mathcal{M}^N_n$ are
one-dimensional, and so the additional tensor product structure
within the irreps is not required; for a general superselection
rule, they can be non-trivial.  The subsystems $\mathcal{N}^A_n$ are
$G$-invariant noiseless subsystems relative to the decoherence map
$\mathcal{G}_A$~\cite{Kni00}. The action of $\mathcal{G}_A$ on a
density operator $\rho$ in terms of this decomposition is
\begin{equation}
  \mathcal{G}_A[\rho] = \sum_n \mathcal{D}_{An} \otimes
  \mathcal{I}_{An}(\Pi^A_n \rho \Pi^A_n )\,,
\end{equation}
where $\Pi^A_n$ is the projection onto the charge sector $n$,
$\mathcal{D}_{An}$ is the trace-preserving map that takes every
operator on $\mathcal{M}^A_n$ to a constant times the identity
operator on that space, and $\mathcal{I}_{An}$ is the identity map
over operators in the space $\mathcal{N}^A_n$.  The effect of the
local superselection rule, then, is to remove the ability to prepare
states or measure operators that have coherence between different
local charge sectors or that are not completely mixed over the
subsystems $\mathcal{M}^A_n$.  The same structure arises for
$\mathcal{H}^B$ and provides an analogous decomposition of
$\mathcal{G}_B$. For further details, see~\cite{BW03,KMP04}.

To address the issue of distillability of a state, we now
demonstrate how to treat multiple systems under a local
superselection rule. If the system that Alice exchanges with Charlie
is made up of several systems, $\mathcal{H}^A=\bigotimes_i
\mathcal{H}^{A_i}$, which are all defined relative to Alice's frame,
the uncertainty in the element $g\in G$ relating Alice's frame to
Charlie's is represented by Eq.~(\ref{eq:AveragedState}) using the
tensor representation $T^A = \bigotimes_i T^{A_i}$.

\subsection{The analogy: general results}

We now present our main results which demonstrate that the structure
of mixed-state entanglement is analogous in many respects to the
structure of pure-state entanglement with a general restriction on
local operations.

The set of LOCC operations that are locally $G$-invariant will be
denoted by $G$-LOCC.  The set of pure bi-partite states that are
locally preparable under a superselection rule for $G$, that is,
preparable by $G$-LOCC, will be denoted by LP$_{G\text{-SSR}}$. A
pure bi-partite state is in LP$_{G\text{-SSR}}$ iff (i) the state is
a product state, and (ii) it is locally $G$-invariant. (Thus, not
all pure product states are in LP$_{G\text{-SSR}}$.)  A state
$|\psi\rangle$ is 1-distillable with $G$-LOCC, denoted
1-D$_{G\text{-SSR}}$, if there exists an operation $\mathcal{E}$ in
$G$-LOCC mapping $|\psi\rangle$ onto a $2{\times}2$-dimensional
space such that $\mathcal{E}[|\psi\rangle\langle\psi |]$ is locally
$G$-invariant and non-separable.  It follows from the main theorem
of Bartlett and Wiseman~\cite{BW03} that $|\psi\rangle$ is in
1-D$_{G\text{-SSR}}$ iff $\mathcal{G}_A \otimes
\mathcal{G}_B[|\psi\rangle\langle\psi |]$ is 1-distillable with
unrestricted LOCC.  Both LP$_{G\text{-SSR}}$ and
1-D$_{G\text{-SSR}}$ are non-empty; explicit examples of each can be
constructed as product/non-product states within 2$\times$2
subspaces or subsystems that are invariant relative to
$\mathcal{G}_A \otimes \mathcal{G}_B$.

\begin{result}
With LOCC constrained by a local superselection rule for $G$, the
classes of pure bi-partite states that are locally preparable
\emph{(LP$_{G\text{-SSR}}$)} or 1-distillable
\emph{(1-D$_{G\text{-SSR}}$)} are both nonempty.
\end{result}

As with mixed-state entanglement, there is a proper gap between
these two classes.  The class of states in the gap contains both
product and non-product pure states, and is analogous to the class
of 1-bound states in mixed-state entanglement. An explicit example
of such a state is a product state that is not locally $G$-invariant
for one or both parties.

\begin{result}
With LOCC constrained by a local superselection rule for $G$, there
exists a non-empty class of states that are neither locally
preparable nor 1-distillable (neither in \emph{LP$_{G\text{-SSR}}$}
nor in \emph{1-D$_{G\text{-SSR}}$}).
\end{result}

Moreover, it is possible to extend $G$-LOCC in such a way that any
pure state in this gap becomes either locally preparable or
1-distillable.  One simply lifts the restriction of the local
superselection by providing Alice and Bob with Charlie's local
frame, so that the local frames of the three parties are
correlated. With this additional resource, Alice and Bob can now
implement any operation in LOCC.  Extending $G$-LOCC to LOCC
divides the proper gap between LP$_{G\text{-SSR}}$ and
1-D$_{G\text{-SSR}}$ into two classes, both of which are
non-empty.  All product states that are not locally $G$-invariant
(i.e., product states not in LP$_{G\text{-SSR}}$) \emph{become}
locally preparable with $G$-LOCC given the shared reference frame
for $G$.  We denote this class BLP$_{G\text{-SSR}}$.  This result
follows directly from the fact that all pure product states are
locally preparable with unrestricted LOCC.  All non-product states
$|\psi\rangle$ for which $\mathcal{G}_A \otimes
\mathcal{G}_B[|\psi\rangle\langle\psi |]$ is not 1-D (i.e.,
non-product states not in 1-D$_{G\text{-SSR}}$) \emph{become}
1-distillable with $G$-LOCC given the shared reference frame for
$G$.  We denote this class B1-D$_{G\text{-SSR}}$.  This result
follows directly from the fact that all pure non-product states
are 1-distillable with unrestricted LOCC.

\begin{result}
With LOCC constrained by a local superselection rule for $G$ and
the additional resource of a shared local reference frame for $G$,
the superselection rule is ``lifted,'' and all states in the
proper gap either become locally preparable
\emph{(BLP$_{G\text{-SSR}}$)} or become 1-distillable
\emph{(B1-D$_{G\text{-SSR}}$)}.  Both classes
\emph{BLP$_{G\text{-SSR}}$} and \emph{B1-D$_{G\text{-SSR}}$} are
nonempty.
\end{result}

Thus, we have demonstrated that the structure of Fig.~\ref{fig:Fig1}
for mixed-state entanglement is analogous to the structure of
pure-state entanglement under the restriction of a superselection
rule.

Although it is likely that the processes of activation and
multi-copy distillation also exist for general superselection rules,
we only consider this aspect of the analogy in depth in the context
of Abelian superselection rules
We turn to this in the next section.

\section{Activation and distillation of pure states constrained by
an Abelian superselection rule}

Although it has proven difficult to fully characterize activation
and distillation processes in the context of mixed-state
entanglement, it is straightforward to do so in the context of pure
states with an Abelian superselection rule, as we now demonstrate.
In particular, we completely classify all pure bi-partite states in
terms of the number of copies needed for distillation.

An Abelian superselection rule is a superselection rule for the
group $H$, all the elements of which commute.  (In the following,
$H$ refers exclusively to an Abelian group.)  Superselection rules
for local charge or particle number are examples, with the relevant
Abelian group being U(1). The superselection rule for photon number
considered in Sec.~\ref{sec:optics} is another example, which can be
seen from the fact that the phase degree of freedom in quantum
optics transforms via the U(1) group, so that a shared phase
reference is an example of a shared reference frame for U(1).  In
the following, we will refer to the superselected quantity for a
Abelian superselection rule as a ``charge'', and we will refer to
local charge eigenstates simply as eigenstates.  We begin with a
useful lemma.  (Note that this lemma fails for the case of
non-Abelian groups.)

\begin{lemma} If $|\Psi\rangle \in \mathcal{H}^{A}\otimes
\mathcal{H}^{B}$ is a non-product state, then Alice and Bob can,
with \emph{$H$-LOCC}, project $|\Psi\rangle$ onto a $2{\times}2$
subspace $\mathcal{S}^{A}\otimes \mathcal{S}^{B}$ with local
projectors $\Pi^A$ and $\Pi^B$, such that $(\Pi^A \otimes
\Pi^B)|\Psi\rangle$ is a non-product state.
\end{lemma}

\begin{proof}
Express $\left\vert \Psi \right\rangle$ using an
eigenstate basis $\{|n,\alpha\rangle_A \}$ for $\mathcal{H}^{A}$ as
\begin{equation}
  \left\vert \Psi \right\rangle =\sum_{n,\alpha} \left\vert n,\alpha
  \right\rangle_A \otimes \left\vert \phi _{n,\alpha
  }\right\rangle_B\,,
\end{equation}
where $n$ labels the ``charge'' and $\alpha$ other quantum
numbers. (The states $|\phi_{n,\alpha}\rangle_B$ are not
necessarily orthogonal and are not normalized.)
For any non-product state $|\Psi\rangle$ there must exist at least
two noncolinear $\left\vert \phi _{n,\alpha}\right\rangle_B$ in this
decomposition, and thus there exists a two-dimensional subspace
wherein the projections of these two vectors are noncolinear.
\end{proof}

From this Lemma, it follows that a state $|\Psi\rangle$ is in
1-D$_{H\text{-SSR}}$ iff there exists a subspace
$\mathcal{S}^{A}\otimes \mathcal{S}^{B}$ that is locally
$H$-invariant such that $(\Pi^A \otimes \Pi^B)|\Psi\rangle \neq 0$.
It follows that for a non-product state that is not in
1-D$_{H\text{-SSR}}$, that is, a state in B1-D$_{H\text{-SSR}}$, any
subspace $\mathcal{S}^{A}\otimes \mathcal{S}^{B}$ such that $(\Pi^A
\otimes \Pi^B)|\Psi\rangle$ is a non-product state must fail to be
locally $H$-invariant.

\subsection{Activation under a local Abelian superselection rule}

\begin{thmnonum}[activation]
For all $|\Psi \rangle \in$ \emph{B1-D$_{H\text{-SSR}}$}, there
exists a $\left\vert \chi \right\rangle \in$
\emph{BLP$_{H\text{-SSR}}$} such that $ \left\vert \Psi
\right\rangle \otimes \left\vert \chi \right\rangle$ is in
\emph{1-D$_{H\text{-SSR}}$}.  We say that $\left\vert \chi
\right\rangle $ has \emph{activated} the entanglement in $\left\vert
\Psi \right\rangle$.
\end{thmnonum}

\begin{proof}
Let
\begin{equation}
  \label{eq:act1}
    |\psi\rangle \equiv (\Pi^{A_1} \otimes \Pi^{B_1})|\Psi\rangle
    \,,
\end{equation}
be a non-product state on a $2\times2$ subspace
$\mathcal{S}^{A_1}\otimes \mathcal{S}^{B_1}$.  Let $\{
|\tilde{n}\rangle_{A_1}, |\tilde{n}'\rangle_{A_1}\}$ be a basis of
eigenstates for $\mathcal{S}^{A_1}$, where
$\tilde{n}\equiv(n,\alpha)$ and $\tilde{n}'\equiv(n',\alpha')$; note
that it can occur that $n=n'$ due to the existence of other quantum
numbers $\alpha$.  Similarly, let $\{ |\tilde{m}\rangle_{B_1},
|\tilde{m}'\rangle_{B_1}\}$ form a basis for $\mathcal{S}^{B_1}$.
Because $|\Psi\rangle$ is in B1-D$_{H\text{-SSR}}$,
$\mathcal{S}^{A_1}\otimes \mathcal{S}^{B_1}$ must fail to be
$H$-invariant, and therefore either $n \ne n'$ or $m \ne m'$ or
both. Let $\mathcal{S}^{A_2}\otimes \mathcal{S}^{B_2}$ be defined
analogously to $\mathcal{S}^{A_1}\otimes \mathcal{S}^{B_1}$, and
define a state $|\chi\rangle$, confined to these subspaces, as
follows:
\begin{equation}
  |\chi\rangle \equiv (|\tilde{n}\rangle_{A_2} +
  |\tilde{n}'\rangle_{A_2}) \otimes (|\tilde{m}\rangle_{B_2} +
  |\tilde{m}'\rangle_{B_2})\,.
\end{equation}
Because either $n \ne n'$ or $m \ne m'$ or both, this state is not
locally $H$-invariant, and therefore is in BLP.  Let
$\mathcal{S}^A_{n+n'}$ be the $H$-invariant subspace of
$\mathcal{S}^{A_1} \otimes \mathcal{S}^{A_2}$ spanned by
\begin{equation}
  \{ |\tilde{n}\rangle_{A_1}\otimes|\tilde{n}'\rangle_{A_2},
  |\tilde{n}'\rangle_{A_1}\otimes|\tilde{n}\rangle_{A_2}\}\,.
\end{equation}
Define $\mathcal{S}^B_{m+m'}$ similarly.  Projecting
$|\psi\rangle\otimes |\chi\rangle$ onto $\mathcal{S}^A_{n+n'}
\otimes \mathcal{S}^B_{m+m'}$ can be performed probabilistically
with $H$-LOCC, and can easily be shown to result in a non-product
state that is locally $H$-invariant.
\end{proof}

\subsection{Distillation under a local Abelian superselection rule}

We now present a complete characterization of the distillability
properties of any pure state constrained by a local Abelian
superselection rule. Specifically, we present two protocols for
distillation of $|\Psi\rangle \in $ B1-D$_{H\text{-SSR}}$. Protocol
A requires three copies of the state, and is based on activation.
Protocol B requires only two copies of the state.  In both protocols
one chooses a local $2\times 2$ subspace $\mathcal{S}^{A_1}\otimes
\mathcal{S}^{B_1}$ with certain properties, and $|\psi\rangle$ is
defined in terms of $|\Psi\rangle$ as in Eq.~(\ref{eq:act1}).
\medskip

\noindent\textit{Distillation Protocol A:} This protocol works if
$\mathcal{S}^{A_1} \otimes \mathcal{S}^{B_1}$ can be chosen such
that ${}_{A_1}\langle \tilde{n} |\psi\rangle \in \mathcal{S}^{B_1}$
or ${}_{A_1}\langle \tilde{n}' |\psi\rangle \in \mathcal{S}^{B_1}$
is \emph{not} locally $H$-invariant, and ${}_{B_1}\langle
\tilde{m}|\psi\rangle \in \mathcal{S}^{A_1}$ or ${}_{B_1}\langle
\tilde{m}'|\psi\rangle \in \mathcal{S}^{A_1}$ is not locally
$H$-invariant (this requires $n \ne n'$ and $m \ne m'$). Projecting
each copy $k$ onto $\mathcal{S}^{A_k} \otimes \mathcal{S}^{B_k}$
yields, with some probability, three copies of $|\psi\rangle$. On
the first copy, Alice measures
$\{|\tilde{n}\rangle_{A_1},|\tilde{n}'\rangle_{A_1}\}$, and on the
second copy Bob measures
$\{|\tilde{m}\rangle_{B_2},|\tilde{m}'\rangle_{B_2}\}$ thereby
collapsing $A_2$ and $B_1$, with some probability, to a product
state that is not locally $H$-invariant, and thus in
BLP$_{H\text{-SSR}}$. It can be shown, by following the proof of the
activation theorem, that this product state is in fact sufficient to
activate the entanglement in the third copy of $|\psi\rangle$. In
this case, $|\Psi\rangle$ is in 3-D$_{H\text{-SSR}}$.  \medskip

\noindent\textit{Distillation Protocol B:}  Consider two copies of
$|\Psi\rangle$. The protocol requires one to project the first
copy onto $\mathcal{S}^{A_1} \otimes \mathcal{S}^{B_1}$ and the
second copy onto $\mathcal{S}^{A_2} \otimes \mathcal{S}^{B_2}$,
and then to project both copies onto $\mathcal{S}^A_{n+n'} \otimes
\mathcal{S}^B_{m+m'}$ (this subspace is defined in the proof of
the activation theorem). The resulting state has the form
\begin{equation}
  \lambda_+|S_+\rangle_A|S_+\rangle_B +
  \lambda_-|S_-\rangle_A|S_-\rangle_B\,,
\end{equation}
where
\begin{equation}
  |S_\pm\rangle_A = |\tilde{n}\rangle_{A_1}|\tilde{n}'\rangle_{A_2}\pm
  |\tilde{n}'\rangle_{A_1}|\tilde{n}\rangle_{A_2}\,,
\end{equation}
and similarly for $|S_\pm\rangle_B$.  It can be shown that
$\lambda_-$ is necessarily non-zero. Thus, if $\lambda_+ \neq 0$,
then the resulting state is a non-product locally $H$-invariant
state, and $|\Psi\rangle$ is in 2-D$_{H\text{-SSR}}$.\medskip

\begin{thmnonum}[distillation]
All pure non-product states are distillable using $H$-LOCC, with at
most three copies of the state required for distillation, that is,
B1-D$_{H\text{-SSR}}\ \cup$ 1-D$_{H\text{-SSR}} =$
3-D$_{H\text{-SSR}}$.
\end{thmnonum}

\begin{proof}
For every state $|\Psi\rangle$ in B1-D$_{H\text{-SSR}}$ for which
Distillation Protocol A fails, Protocol B necessarily succeeds. The
proof is as follows. If Protocol A fails, then for all choices of
$\mathcal{S}^A \otimes \mathcal{S}^B$, the state $|\psi\rangle$ has
a Schmidt basis~\cite{Nie00} composed of local eigenstates, and
Protocol B can be shown to work whenever $|\psi\rangle$ is of this
form.
\end{proof}

Although related, this theorem is different from the one presented
in Schuch \emph{et al}~\cite{Sch04b}.  In~\cite{Sch04b}, it was
shown how states constrained by an Abelian superselection rule can
be converted, using at most three copies, to the state $|{\rm
V{-}EPR}\rangle = |0\rangle_A|1\rangle_B + |1\rangle_A|0\rangle_B$.
As we argued in Sec.~\ref{sec:optics}, this state can be considered
to be bound entangled when local operations are constrained by a
photon-number superselection rule. In addition, it is shown
in~\cite{Sch04b} that this resource can be asymptotically converted
to non-product locally $H$-invariant states.  Our result is stronger
in that we show it is possible to prepare (with some probability) an
effective two-qubit entangled state that is \emph{locally
$H$-invariant} using only operations obeying the superselection rule
and at most three copies of any entangled pure state.  This
effective two-qubit state may then be distilled by standard
techniques to prepare locally $H$-invariant maximally entangled pure
states at some asymptotic rate.

Finally, we note that we can completely characterize the
distillability properties of any pure state constrained by a local
Abelian superselection rule.  We show that the class B1-D can be
divided into three non-empty regions by establishing that
1-D$_{H\text{-SSR}}$ is a \emph{proper} subset of
2-D$_{H\text{-SSR}}$ and 2-D$_{H\text{-SSR}}$ is a \emph{proper}
subset of 3-D$_{H\text{-SSR}}$ (i.e. Protocol A sometimes fails
while Protocol B succeeds). States such as
\begin{align}
  |\psi_{\text{2-D}}'\rangle &= \tfrac{1}{\sqrt{2}}
  (|01\rangle_A |0\rangle_B + |10\rangle_A |1\rangle_B) \,, \\
  |\psi_{\text{2-D}}''\rangle &= \tfrac{1}{\sqrt{2}}
  (|01\rangle_A |{+}\rangle_B +|10\rangle_A |{-}\rangle_B) \,, \\
  |\psi_{\text{2-D}}'''\rangle &= \tfrac{1}{\sqrt{2}}
  (|0\rangle_A |1\rangle_B + |1\rangle_A |0\rangle_B) \,,
\end{align}
expressed in the Fock basis, are in 2-D$_{H\text{-SSR}}$ (using
Protocol B). None of them are locally $H$-invariant and therefore
are not in 1-D$_{H\text{-SSR}}$. The state
\begin{equation}
  |\psi_{\text{3-D}}\rangle = \tfrac{1}{\sqrt{2}}
  (|0\rangle_A |{+}\rangle_B +|1\rangle_A |{-}\rangle_B) \,,
\end{equation}
is in 3-D$_{H\text{-SSR}}$ (as is any state in B1-D).  However,
because two copies of this state become separable when averaged
(uniformly) over $H$ locally, it is not in 2-D$_{H\text{-SSR}}$.

\section{Discussion}

In summary, we have shown how to reproduce the rich classification
scheme of mixed-state entanglement by restricting local operations
on the set of pure states so as to create a proper gap between what
is locally preparable and what is 1-distillable. Debates over the
entanglement properties of pure states under restricted operations,
such as have appeared in the quantum optics literature, are resolved
by recognizing novel categories of entanglement in this context. Our
results suggest that the exotic structure of mixed-state
entanglement is generic, and that developing entanglement theory
under other sorts of restrictions is a promising direction for
further research.

For example, recent interest in creating bi-partite entangled states
in condensed matter systems requires careful articulation of the
operational meaning of entanglement, due to the various practical
restrictions on operations on these systems.  Local particle-number
superselection rules often apply in practice, and as noted
in~\cite{Wis03,Sam05,Bee05}, for example, the single-electron
two-mode Fock state $\frac{1}{\sqrt{2}}(|0\rangle_A|1\rangle_B +
|1\rangle_A |0\rangle_B)$ has ambiguous entanglement properties
under this restriction.  Wiseman and Vaccaro~\cite{Wis03} have
introduced an operational measure, called \emph{entanglement of
particles} to quantify the distillable entanglement under a local
particle-number superselection rule, and this two-mode
single-electron Fock state has no entanglement of particles by this
measure.  For this reason, most proposals for creating bi-partite
entangled states make use of spin or orbital angular momentum
degrees of freedom of multiple particles~\cite{Sam03,Bee03,Sam04}.
We note, however, that the two-mode single-electron Fock state is an
entanglement resource akin to the two-mode single-photon state,
which we have shown to be useful through activation or multi-copy
distillation; also, a suitable shared U(1) reference frame could
``lift'' the restriction of the superselection rule, and the
two-mode single-electron Fock state would be unambiguously entangled
with such a resource. Moreover, entangled states between angular
momentum degrees of freedom of different particles will yield no
real advantage over the two-mode single-electron Fock state in
situations wherein there is a local SU(2) superselection rule. Such
a superselection rule will be in force, for instance, if the parties
fail to share a Cartesian frame for spatial
orientations~\cite{Bar03}.  As with quantum optical systems, we
emphasise the need to be operational when classifying or quantifying
entanglement.

The theory of entanglement for indistinguishable particles is
another situation where our results may shed some light.  Recent
research has investigated the ``quantum correlation between
particles''~\cite{Pas01,Sch01}, which relates to the correlations
between indistinguishable particles inherent in the symmetry (or
antisymmetry) of a many-particle wavefunction.  Refs.~\cite{Wis03}
argues that these quantum correlations are merely ``fluffy bunny
entanglement''~\cite{Wis03b}, that is, operationally useless.  Our
work here supports this conclusion; we would say that the
entanglement is bound by the restriction of the indistinguishability
of particles.  Nonetheless, in analogy with restrictions arising
from superselection rules, it may be worthwhile to consider the
possibility of ``lifting'' this restriction through an appropriate
shared resource.

The analogy we present here also suggests that it may be fruitful to
think of standard LOCC as a restriction relative to the ``more
natural'' PPT-preserving operations, and to consider whether a
resource that lifts this restriction might be established with the
same ease as a shared reference frame.

\begin{acknowledgments}
This work was supported by the Australian Research Council and the
State of Queensland.  We thank John Vaccaro for many fruitful
discussions and results on the entanglement properties of a single
photon. We also thank Konrad Banaszek, Lucien Hardy, Michael
Nielsen, Terry Rudolph, Steven van Enk and Andrew White for helpful
discussions.
\end{acknowledgments}

\end{document}